\documentclass[12pt]{article}
\pdfoutput=1

\usepackage[T1]{fontenc}
\usepackage{amsmath,amssymb,amsfonts}
\usepackage{cite}
\usepackage{subcaption}
\usepackage{graphicx}
\usepackage{hyperref}
\usepackage{microtype}
\usepackage{color}
\usepackage{multirow}

\hypersetup{
colorlinks=true, linkcolor=blue, citecolor=blue, filecolor=blue, urlcolor=blue }

\renewcommand{\arraystretch}{1.25}

\newcommand{\citere}[1]{Ref.~\cite{#1}}
\newcommand{\citeres}[1]{Refs.~\cite{#1}}
\newcommand{\refsec}[1]{Section~\ref{#1}}
\newcommand{\refapp}[1]{Appendix~\ref{#1}}
\newcommand{\reffig}[1]{Figure~\ref{#1}}
\newcommand{\reftab}[1]{Table~\ref{#1}}
\newcommand{\refeq}[1]{(\ref{#1})}

\newcommand{\pldist[1]}{\left[\frac{1}{1- #1} \right]_+}
\newcommand{\pllogdist[1]}{\left[\frac{\log (1-z)}{1 - #1} \right]_+}

\begin{document}

\begin{titlepage}

\begin{flushright}
{\small
TTK-19-10\\
MS-TP-19-06\\
ZU-TH 12/19\\
P3H-19-004\\[2mm]
\today
}
\end{flushright}

\vskip1cm
\begin{center}
{\Large \bf
Threshold Resummation for Dark-Matter Production\\[0.2cm] 
at the LHC}
\end{center}

\vspace{0.8cm}
\begin{center}
{\sc M.~Kr\"amer${}^a$, A.~Kulesza${}^b$, A.~M\"uck${}^a$, R.~Sch\"urmann${}^{a,c}$} \\[6mm]
{
${}^a$ Institut f\"ur Theoretische Teilchenphysik und Kosmologie,
RWTH Aachen University,\\
D-52056 Aachen, Germany\\
${}^b$ Institute for Theoretical Physics, WWU M\"unster,\\ D-48149 M\"unster, Germany\\
${}^c$ Physik-Institut, Universit\"at Z\"urich, Wintherturerstrasse 190,\\ CH-8057 Z\"urich, Switzerland}
\\[0.3cm]
\end{center}

\vspace{0.7cm}
\begin{abstract}
\vskip0.2cm\noindent

We derive precision predictions for the production of dark-matter particles recoiling against a jet with large transverse 
momentum at the LHC. The dark-matter fermions are described within a simplified model and couple to the Standard Model 
via a vector mediator. Our predictions for the mono-jet signature include the resummation of the leading and next-to-leading 
threshold logarithms. The corresponding matching coefficient is evaluated at NLO. 
The resummed result is matched to the fixed-order NLO cross section obtained from the MadGraph framework. 
We discuss numerical results for several benchmark scenarios at the LHC.

\end{abstract}
\end{titlepage}

\section{Introduction}
\label{sec:Introduction}

The nature of dark matter remains one of the big unanswered questions in modern physics. Weakly interacting massive particles (WIMPs) are well-motivated dark matter candidates and are predicted by many extensions of the Standard Model (SM)~\cite{PDG}.
One of the main goals of the Large Hadron Collider (LHC) is to test the WIMP paradigm, see e.g.~\citere{Kahlhoefer:2017dnp}. 
Many LHC searches for WIMP dark matter rely on the mono-X signature, where one would observe a SM particle recoiling against the missing transverse energy caused by WIMPs escaping the detector. 

To interpret the LHC data, precise predictions for the dark-matter production cross section and for the
corresponding SM backgrounds are essential. The main irreducible SM background is Z-boson
production in association with jets, where the Z boson decays invisibly into neutrinos. A thorough discussion of the
theoretical description of this dark-matter background can be found in \citere{Lindert:2017olm}. 

Predictions for the dark-matter signal depend on the concrete model under consideration.
We employ the simplified-model approach, \citere{Buchmueller:2013dya,Abdallah:2014hon}, and
derive precision predictions for the process $pp \to \chi \bar{\chi} + \text{jet}$ in a simplified dark-matter
model with a fermionic dark-matter particle $\chi$ and an s-channel vector mediator $Y$ which couples to SM quarks but not to
leptons~\cite{Abdallah:2015ter}. Hence, stringent bounds from resonance searches in the Drell-Yan process are avoided. The LHC has also
placed bounds on the simplified model parameter space employing resonance searches in the di-jet final 
state~\cite{Sirunyan:2018xlo,ATLAS-2018-51,Fairbairn:2016iuf}. 

Mono-X production of dark matter has already been investigated in simplified models at next-to-leading order (NLO) in 
QCD~\cite{Fox:2012ru,Haisch:2013ata,Abercrombie:2015wmb,Backovic:2015soa,Neubert:2015fka,Das:2016pbk,Kraml:2017atm,Afik:2018rxl}.
The simplified dark-matter model with a vector mediator has been analyzed at NLO
including jet merging and parton-shower effects~\cite{Backovic:2015soa}.
The model has been implemented in the MadGraph5\_aMC@NLO-framework~\cite{Alwall:2014hca} using
FeynRules~\cite{Alloul:2013bka,Degrande:2014vpa} to obtain NLO predictions in a 
fully automated way. The NLO corrections result in a moderate enhancement of the integrated
cross section and the missing transverse momentum distribution, depending in detail on the mediator and dark-matter masses. Moreover, a sizeable reduction of the scale dependence of the cross section is 
reported improving the predictions from LO to NLO, as expected. 
Given the current WIMP exclusion limits from the LHC~\cite{Sirunyan:2018xlo,ATLAS-2018-51} and from direct-detection experiments, see e.g.~\citere{Aprile:2018dbl}, future LHC searches will focus on large dark-matter and/or mediator masses.  In this case, a significant part of the cross section can be attributed to the threshold region, where large logarithmic corrections from soft and collinear parton emission beyond NLO should be taken into account. 

In this work, we refine the theoretical prediction of \citere{Backovic:2015soa} for inclusive observables in the process 
$pp \to \chi \bar{\chi} + \text{jet}$  by resumming potentially large threshold logarithms up to next-to-leading logarithmic (NLL) accuracy.  More specifically, we perform the resummation for the double-differential 
cross section with respect to the transverse momentum and the invariant mass of the dark-matter particles in the
final state. To calculate the anomalous dimensions for the renormalization-group evolution of the soft and the jet functions
as well as the soft and collinear NLO contributions to the matching coefficient, we use the 
strategy of regions \cite{Beneke:1997zp}. The NLL expressions are identical 
to those presented in~\citere{deFlorian:2005fzc} after adapting the color-factors. 
The corresponding matching coefficient is calculated at NLO in the strong coupling $\alpha_s$. We call these predictions NLL' to distinguish them 
from NLL predictions with a LO matching coefficient. The hard function, which only enters the
matching coefficient, is obtained numerically from MadLoop within the 
MadGraph5\_aMC@NLO-framework~\cite{Alwall:2014hca}. Combining these numerical one-loop results with the
analytic results obtained using the strategy of regions provides a convenient and straightforward 
way to calculate NLL' resummed cross sections. Since the invariant mass of the dark-matter particles is not accessible experimentally, we integrate the resummed distribution over the invariant mass. 
The resummed predictions are matched to the full fixed-order NLO prediction obtained from the MadGraph5\_aMC@NLO-framework~\cite{Alwall:2014hca}.

The resummation of threshold logarithms has been achieved for closely related SM processes. In particular, threshold resummation in the direct QCD approach was derived to NLL accuracy for direct photon hadroproduction  in~\citeres{Catani:1998tm, Catani:1999hs, Laenen:1998qw, Sterman:2000pt, deFlorian:2005wf }. The resummed results in one-particle inclusive kinematics were later used to obtain approximate 
higher-order corrections for direct photon hadroproduction as well as for the production of a massive gauge boson at high transverse momentum \cite{Kidonakis:1999hq, Kidonakis:1999ur, Kidonakis:2003xm,Gonsalves:2005ng}. A general discussion of  threshold resummation for 
single inclusive cross sections at NLL and NNLL can be found in \citeres{Laenen:1998qw, Catani:2013vaa, Hinderer:2018nkb}. For the hadroproduction of Higgs bosons at large transverse momentum, threshold 
resummation is known to NLL' accuracy \cite{deFlorian:2005fzc}. Recently, a similar analysis for the closely related Drell-Yan process at a fixed invariant mass and high transverse momentum has been performed in~\citere{Bacchetta:2019tcu}.
Within the  formalism  of Soft-Collinear-Effective-Field-Theory (SCET)  threshold resummation to NNLL
accuracy has been achieved in  
\citere{Becher:2009th} for direct photon production,  in \citere{Becher:2011fc}  for the production of $W, Z$ bosons, and in  \citere{Huang:2014mca} for Higgs-boson production at high transverse momentum. 

This work is organized as follows: In \refsec{sec:model}, we briefly introduce the simplified dark-matter model. 
The kinematics and the partonic channels for dark-matter production in association with a jet are discussed in
\refsec{sec:partonic}.
The resummation of the threshold logarithms is discussed in \refsec{sec:resummed} where we derive analytical results
for the soft and the jet function and calculate the hard function numerically.
In \refsec{sec:results}, numerical results for dark-matter production at the LHC are
presented for several benchmark scenarios and the effects of resummation are investigated. 
Finally, we conclude in \refsec{sec:Conclusions} and present the analytic LO cross sections in \refapp{app:LO}.

\section{Dark-Matter Model}
\label{sec:model}

The simplified dark-matter model which we investigate in this work consists of a
massive s-channel vector mediator $Y$ and a Dirac dark-matter fermion $\chi$. 
The interaction part of the Lagrangian in
the simplified dark matter model reads
\begin{equation} 
\mathcal{L}_{\text{int.}} \supset \bar{\chi} \gamma_{\mu} \left( g^V_{\chi} + g^A_{\chi} \gamma^5 \right) \chi Y^{\mu} + \bar{q} \gamma_{\mu} \left(g^V_{SM} + g^A_{SM} \gamma^5 \right) q Y^{\mu}.
\end{equation}
The first term in the Lagrangian describes the interaction of the mediator with the
dark-matter fermions. We allow for a vector- and an axial-vector coupling. The second term
describes the interaction of the mediator with the quark fields $q$. Again we allow
for a vector- and an axial-vector coupling. In addition, we assume that the
couplings are flavour diagonal so that no flavour mixing is introduced. Consequently,
our simplified model has six free parameters, i.e.
\[
m_Y \, , \, m_{\chi} \, , \, g^V_{\chi} \, , \, g^A_{\chi} \, , \, g^V_{SM} \, , \, g^A_{SM}
\, .
\]
The benchmark points considered in the numerical analysis are given by the fixed
couplings
\begin{equation}
g^V_{\chi}=1 \, , \, g^A_{\chi}=0 \, , \, g^V_{SM} = 0.25 \, , \, g^A_{SM} =0
\end{equation}
and we vary the masses of the mediator and the dark-matter particle. We stick to these 
values for the couplings which are often used as benchmark points in the literature although
large regions of parameter space for the mediator and dark-matter masses are already 
excluded~\cite{Sirunyan:2018xlo,ATLAS-2018-51}. However, viable models can be obtained 
by choosing smaller couplings, e.g.\ $g^V_{SM}=0.1$ or $g^A_{SM}=0.1$. As discussed in \refsec{sec:results}, the exact choice for
the couplings does not have a relevant impact on the relative size of the NLO or matched NLO+NLL' corrections. 

\section{Partonic Channels and Kinematic Considerations}
\label{sec:partonic}

We calculate predictions for the production of a pair of dark-matter fermions $\chi$
recoiling against a jet $j$. At LO, the corresponding partonic processes are
\begin{equation}
\begin{aligned}
q\bar{q} & \to \chi \bar{\chi} + g \, , \\
qg & \to \chi \bar{\chi} + q \, , \\
\bar{q}g & \to \chi \bar{\chi} + \bar{q} \, ,
\end{aligned}
\end{equation}
where $g$ denotes a gluon, $q$ a quark, and $\bar{q}$ an anti-quark. 
In \refsec{sec:resummed}, we label the different contributions to the NLL'
cross section by the final-state parton at LO,
e.g.\ the cross section $\mathrm{d}\sigma_g$ refers to the first process.

The central quantity for threshold resummation is the double
differential cross section with respect to the transverse momentum $p_T$ of 
the pair of the dark-matter particles in the final state and
their invariant mass $M^2$, i.e.\
\begin{equation}
\frac{\text{d}^2\sigma(s)}{\text{d}p_T^2 \text{d} M^2} = \sum_{i,j}
\int_0^1 \text{d} x_1 \, f_{i}(x_1) \int_0^1 
\text{d} x_2 \, f_{j}(x_2) 
\frac{\text{d}^2 \hat{\sigma}_{ij}(\hat{s})}{\text{d} p_T^2 \text{d}M^2},
\label{eq:fac_hadronic_cross}
\end{equation}
where the squared partonic center-of-mass energy is given by $\hat{s}=x_1x_2 s$
in terms of the squared collider energy $s$. The parton distribution functions 
(PDFs) are denoted by $f_i$ and the sums run over all 
contributing initial-state partons $i,j$.

The threshold energy $\sqrt{\hat{s}_T}$ to produce a final state with a given 
$p_T$ and $M^2$ is given by
\begin{equation}
\sqrt{\hat{s}_T} = p_T + m_T = p_T + \sqrt{p_T^2 + M^2}
\end{equation}
and the corresponding partonic threshold variable is defined by
\begin{equation}
\hat{y}_T = \frac{p_T + m_T}{\sqrt{\hat{s}}}.
\end{equation}
The LO cross section can be written as a function of the squared threshold variable 
$\hat{y}_T^2$ and the ratio $r=p_T/m_T$. The explicit results for the LO cross
sections of the different partonic channels are given in \refapp{app:LO}. 
Since the dark-matter particles are not accessible experimentally, 
we have simplified the otherwise fully differential cross section by integrating 
over the dark-matter phase space. Hence, we obtain an effective 
$2 \to 2$ process which, for a given mass $M^2$, has the same kinematics as
$2 \to 2$ SM processes for which threshold resummation has been obtained in the 
past \cite{Becher:2011fc, deFlorian:2005fzc}.

We perform resummation in Mellin space. The Mellin moments are defined as
\begin{equation}
\frac{\text{d}^2 \hat{\sigma}_{ij}(N)}{\text{d} p_T^2 \text{d}M^2} =
\int_0^1 \text{d} \hat{y}_T^2 \, (\hat{y}^2_T)^{N-1} 
\frac{\text{d}^2 \hat{\sigma}_{ij}(\hat{y}_T^2)}{\text{d} p_T^2 \text{d}M^2} \, .
\label{eq:Mellinmoments}
\end{equation} 

We do not require an analytic expression for the full NLO partonic cross 
section in this work. However, it is instructive to inspect
its generic structure written in terms of a threshold expansion
\begin{equation}
\frac{\text{d}^2\hat{\sigma}_{ij}^{\mathrm{NLO}}(\hat{s})}{\text{d} p_T^2 \text{d} M^2} = 
\!\!\int_0^1 \!\!\!\text{d} z \, 
\frac{\text{d}^2 \hat{\sigma}^{\mathrm{LO}}_{ij} (\hat{s}z) }{\text{d}p_T^2 \text{d} M^2}
\bigg( c^{(3)}_{ij} \pllogdist[z] \!\!\!+ c^{(2)}_{ij} \pldist[z] \!\!\!
+ c^{(1)}_{ij} \delta(1-z) \bigg ) + d_{ij}(\hat{y}_T^2)
\label{eq:fac_partonic_NLO}
\end{equation}
with $z = 1 - s_{H}/Q^2$. Here, $s_{H}$ denotes the hadronic mass of the final
state, and $Q^2 = p_T \sqrt{\hat{s}_T}$ is a high-energy scale of this
process which naturally appears in addition to $\hat{s}_T$. For initial-state
parton combinations $i$, $j$ present at LO, the $z$-integration over the plus-distributions results in $\log^2 (1-\hat{y}_T^2)$ and 
$\log (1-\hat{y}_T^2)$ contributions at NLO. These threshold logarithms are
resummed at NLL accuracy to all orders in this work.
The function $d_{ij}$ denotes all terms which vanish in the threshold
limit and are not needed at NLL' accuracy. The 
leading power coefficients of the threshold expansion $c_{ij}^{(3)}$, $c_{ij}^{(2)}$, and parts of 
$c_{ij}^{(1)}$ are derived analytically, 
while the so-called hard contributions to 
$c_{ij}^{(1)}$ are obtained numerically as discussed in \refsec{sec:resummed}. 
Note that
\begin{equation}
Q^2=p_T \sqrt{\hat{s}_T} = \frac{p_T}{p_T+m_T} \hat{s}_T = 
\frac{r}{1+r} \hat{s}_T \, .
\end{equation}
This relation can be used in the soft and jet functions derived in 
\refsec{sec:resummed} to relate the two high-energy scales. As long as $r$ is of 
$\mathcal{O}(1)$, i.e.\ $p_T$ is of the same order as the mass
of the dark-matter system, $\hat{s}_T$ and $Q^2$ are of the same order and
standard threshold resummation holds without the need to take care of an additional
large scale ratio.

As mentioned above, the distribution in $M^2$ is not accessible experimentally. The
transverse momentum distribution or any fiducial cross section is simply 
obtained by integrating the resummed double-differential cross section, 
which is derived in \refsec{sec:resummed}.

\section{Resummed Cross Section}
\label{sec:resummed}

In this section, we outline our strategy to calculate the resummed cross section at
NLL' accuracy. For brevity, the double differential cross section introduced in
\refsec{sec:partonic} is denoted by $\text{d}\sigma$ in the following. 
We use the method of regions to derive all the one-loop
ingredients of the resummation formula. The resummation is performed in Mellin
space.

Using the method of regions, there are the soft, the collinear, and the hard region
which contribute to the cross section near threshold at leading power. To define the different
regions, we decompose the momentum $p$ of emitted radiation into components 
$(n_+ \cdot p,n_- \cdot p,p_\perp)$, where we have introduced light-like 
four-vectors $n_\pm$ with spatial components along and opposite to the jet 
direction ($n_\pm^2=0$, $n_+\cdot n_-=2$). For the soft region $p$ 
scales according to $(\lambda,\lambda,\lambda)$, where the expansion 
parameter $\lambda=1-\hat{y}_T^2$ is small near threshold. 
The collinear and the hard regions scale according to
$(1,\lambda,\sqrt{\lambda})$ and $(1,1,1)$, respectively.

To obtain the resummed cross section
we have to calculate the leading-power result in each region. For the soft and
the collinear region, our calculation leads to simple analytical results. The
contribution of the hard region is obtained numerically as explained below. 

In the soft region, the leading-power expansion of the
squared matrix elements for the real-emission contribution is identical to the
Eikonal approximation for soft-gluon emission. Using the leading-power
expansion also for the phase-space integration, one can write the soft
contribution up to NLO for each partonic channel as a convolution with the 
LO cross section according to
\begin{equation}
\text{d} \hat{\sigma}_{a}^{\mathrm{NLO}, soft} = 
\int_0^{1} \text{d} z \, \text{d} \hat{\sigma}_{a}^{\mathrm{LO}} \left(z \hat{s} \right) 
S^{\mathrm{NLO}}_{a}(z) \, ,
\label{eq:soft_con_dimless}
\end{equation}
where the so-called soft function reads
\begin{equation}
S^{\mathrm{NLO}}_{a}(z,\mu^2) = \delta(1-z) + C_{a} \frac{\alpha_S(\mu^2)}{\pi} \\
\frac{(4\pi)^{\epsilon}}{\Gamma(1-\epsilon)} 
\left( \frac{\mu^{2}}{\hat{s}_T} \right)^{\epsilon} \frac{1}{(1-z)^{1+2\epsilon}} \left[ - \frac{1}{\epsilon} + \mathcal{O}(\epsilon^2) \right].
\label{eq:soft_func_dimless_2}
\end{equation}
The color factors are given by $C_{q}=C_{\bar{q}} = C_F+C_A-C_F = 3$ and 
$C_{g} = 2 C_F - C_A = -1/3$ for the partonic channels, i.e.\ each initial-state quark contributes
$C_F$, an initial-state gluon $C_A$ and for each final-state particle the 
respective color factor is subtracted. Moreover, we work in $d=4-2\varepsilon$ dimensions, and 
$\mu$ is the scale introduced in dimensional regularization
to keep the running coupling dimensionless. In the calculation, we have used that  
the jet momentum is perpendicular to the beam direction at threshold. The term
$(1-z)^{-(1+2\epsilon)}$ can be expanded in $\epsilon$ in terms of a $\delta$-function and
plus-distributions. In the following we consider the
$\overline{\text{MS}}$-renormalized quantity
\begin{equation}
\begin{aligned}
S^{\mathrm{NLO},R}_{a}(z,\mu^2) = \delta(1-z) + \frac{\alpha_S(\mu)}{\pi} C_a
& \left[ \frac{1}{4} \left( \log^2 \frac{\mu^2}{\hat{s}_T} - \frac{\pi^2}{6} \right) \delta(1-z) \right. \\
 &- \left. \log \frac{\mu^2}{\hat{s}_T} \left[ \frac{1}{1-z} \right]_+ + 
 2 \left[\frac{\log (1-z)}{1-z} \right]_+ \right] \, .
\end{aligned}
\label{eq:soft_func_dimles_ren_full}
\end{equation} 
At NLO, the soft functions is, by construction, identical to the SCET soft 
function of~\cite{Becher:2009th}, where a dimensionful convolution variable is used instead.

Since the Mellin transform converts convolutions like \refeq{eq:soft_con_dimless}
into ordinary products, it is beneficial to go to Mellin space. 
The Mellin transform of the renormalized NLO soft function reads
\begin{equation}
S^{\mathrm{NLO}}_{a}(N,\mu^2) = 1 + \frac{\alpha_S(\mu^2)}{\pi} 
C_{a} \left[ \frac{1}{4} \log^2 \frac{\bar{N}^2 \mu^2}{\hat{s}_T} + \frac{\pi^2}{8} 
+ \mathcal{O}\left(\frac{1}{N}\right) \right],
\label{eq:resum_soft_with_Q}
\end{equation}
where we drop all terms suppressed by $1/N$, since power-suppressed terms also
contribute at $\mathcal{O}(1/N)$, and use $\bar{N}=N \exp(\gamma_E)$ with Euler's constant
$\gamma_E$. For the Mellin-space functions we drop the label $R$, since we always consider renormalized
quantities. We
observe that the characteristic scale of the soft function is given by 
$\mu_s^2=\hat{s}_T/\bar{N}^2$. The virtual contributions to the soft function
consist of scaleless integrals and do not contribute.

In analogy, we can consider the collinear region which corresponds to real radiation collinear with the
outgoing parton. The leading-power expansion of the squared matrix-element leads to 
the Born expression multiplied by the relevant splitting function for each
partonic channel. Also expanding the phase space, in analogy to the soft case, one
finds
\begin{equation}
\text{d} \hat{\sigma}_{a}^{\mathrm{NLO},coll} = 
\int_0^{1} \text{d} z \, \text{d} \hat{\sigma}_{a}^{\mathrm{LO}} \left(z \hat{s} \right) 
J_a^{\mathrm{NLO}}(z) \, 
\end{equation}
with the so-called jet function
\begin{equation}
J^{\mathrm{NLO}}_a(z) = \delta(1-z) + \frac{\alpha_S(\mu^2)}{\pi} \\
\frac{(4\pi)^{\epsilon}}{\Gamma(1-\epsilon)} 
\left( \frac{\mu^2}{Q^2} \right)^{\epsilon} \frac{1}{(1-z)^{1+\epsilon}} 
j_a(\epsilon)
\end{equation}
for the outgoing parton $a$, where
\begin{equation}
j_q(\epsilon) = 
\frac{C_F}{2} \left[ - \frac{2}{\epsilon} - \frac{3}{2} + \left( \frac{\pi^2}{3} - \frac{7}{2} \right) \epsilon + \mathcal{O}(\epsilon^2) \right]
\end{equation}
for a quark or an anti-quark 	in the final state and
\begin{equation}
j_g(\epsilon) = 
- \frac{C_A}{\epsilon} - \pi b_0 + \frac{\epsilon}{18}  \left(10 N_f T_R +\frac{C_A}{2} (6 \pi^2 - 67) \right)
+ \mathcal{O}(\epsilon^2)
\end{equation}
for a gluon, which can also split into $N_f=5$ light quark flavours. Here,
$b_0 = \frac{1}{12\pi} (11 C_A - 4 T_R N_f)$ is the first coefficient of the 
QCD $\beta$-function and $T_R=1/2$ for QCD.

In analogy to the soft function, the renormalized NLO jet functions for outgoing partons in Mellin space read
\begin{equation}
J^{\mathrm{NLO}}_q(N,\mu^2) = 1 + \frac{\alpha_S(\mu^2)}{\pi} C_F \left[\frac{1}{2} \log^2 \frac{\bar{N}\mu^2}{Q^2} + 
\frac{3}{4} \log \frac{\bar{N}\mu^2}{Q^2} + \frac{7}{4} - \frac{\pi^2}{6} + \mathcal{O}\left(\frac{1}{N} \right) \right]
\end{equation}
and
\begin{equation}
\begin{aligned}
J^{\mathrm{NLO}}_g(N,\mu^2) = 1 + \frac{\alpha_S(\mu^2)}{\pi} &\left[ \frac{C_A}{2} \log^2 \frac{\bar{N}\mu^2}{Q^2} + \pi b_0 \log \frac{\bar{N}\mu^2}{Q^2} \right. \\
&- \left. \frac{1}{18} \left( 10 N_f T_R + \frac{C_A}{2} (6\pi^2 - 67) \right) + \mathcal{O}\left(\frac{1}{N} \right) \right] .
\end{aligned}
\end{equation}
We observe that the characteristic scale of the jet function is given by 
$\mu_c^2=Q^2/\bar{N}$. The virtual contributions to the jet functions
again consist of scaleless integrals and do not contribute. The calculated jet
functions agree with the SCET jet functions in~\citere{Becher:2009th} at NLO.

Concerning the hard corrections, the real emission of a hard gluon is kinematically
impossible near threshold. In contrast, the hard virtual contribution does not vanish. We
do not attempt to derive or use analytical results for the hard contribution. Instead,
we obtain this contribution numerically from the NLO MadGraph framework as detailed in the following. 
To access the virtual
corrections at threshold, a MadLoop library for the process at hand is most suitable. 
To obtain the one-loop hard function for
given $p_T$ and $M^2$, at leading power we only have to consider phase-space points at threshold. 
As for the LO cross section, we integrate the 
$\overline{\text{MS}}$-renormalized virtual part of the NLO result over
the dark-matter phase-space and divide the result by the respective LO result.
Alternatively, one can use the process with a mediator of mass $M^2$ in the final state.
At threshold, the phase-space is completely fixed up to an azimuthal symmetry and no integration is
needed. One calculates the interference contribution of the $\overline{\text{MS}}$-renormalized 
one-loop and the tree-level amplitude and normalizes by the corresponding squared tree-level amplitude. 

The hard function for given $p_T$ and $M^2$ does neither depend on the dark-matter mass 
nor on the mediator mass. 
It depends only on the variables $Q^2$ and $\hat{s}_T$ representing the hard scale 
$\mu_h$, the scale $\mu$ and 
the renormalization scale $\mu_R$ since the renormalization of the strong coupling is 
part of the virtual corrections. 
Hence, the hard function includes logarithms of the scale ratio $\mu^2/\mu_h^2$.
One can write the hard function as 
a scale independent piece given numerically and the scale-dependent terms which are available
analytically. At the level of the complete matching coefficient, this is 
discussed at the end of the chapter, 
where we choose the hard scale $\mu_h^2=Q^2$ and
the ratio $r=p_T/m_T$ as independent variables instead of $Q^2$ and $\hat{s}_T$.
Since the hard function is a purely virtual correction, it is proportional 
to $\delta(1-z)$ and, therefore, it has no $N$-dependence in Mellin space.

In Mellin space, the threshold limit corresponds to large $N$. Hence, in the
threshold limit, large logarithms can only be avoided in one of the three contributions
(soft, collinear or hard) by a corresponding scale choice but not in all of them. The NLO result therefore necessarily contains large logarithms, no matter
how the scale $\mu$ is chosen. 
Since the factorization scale, where the PDFs are evaluated, is usually taken as equal to the hard scale, we choose to use $\mu \sim \mu_h$.
To control the large logarithms in the soft and the jet function, we use
renormalization-group techniques as introduced in the following. 

It can be shown that the cross section at leading power factorizes to all orders in perturbation theory \cite{Catani:1998tm, Becher:2009th} 
and the Mellin-space cross section is given by an ordinary product of the LO cross section and the soft, the jet and the hard function. Using  SCET convention~\cite{Becher:2009th} one writes 
\begin{equation}
\text{d} \hat{\sigma}_a(N) = S_{a}(N+1) \, J_{a}(N+1) \, H_{a} \,\, \text{d}
\hat{\sigma}^{\mathrm{LO}}_a(N).
\label{eq:resum_final_resumed_cs}
\end{equation}
Note that in this approach the physics of the incoming jets is completely contained in the soft function and only the hard collinear 
emissions of the outgoing jet give rise to the jet function. Hence, the assignment of the different physics contributions to the factorized functions 
differs compared to the direct QCD approach, cf.~\cite{deFlorian:2005fzc}.
The factorized structure implies that the different pieces obey 
renormalization-group equations. For the soft and the jet function, one has
\begin{equation}
\begin{aligned}
\frac{\text{d} \log S_{a}(N,\mu^2)}{\text{d} \log \mu^2} &= 
\gamma^{S_A}_{a}(\alpha_S(\mu^2)) \log \frac{\bar{N}^2\mu^2}{Q^2} - 
\gamma^{S_B}_{a}(\alpha_S(\mu^2)), \\
\frac{\text{d} \log J_a(N,\mu^2)}{\text{d} \log \mu^2} &= 
\gamma^{J_A}_{a} (\alpha_S(\mu^2)) \log \frac{\bar{N}\mu^2}{Q^2} - 
\gamma^{J_B}_{a} (\alpha_S(\mu^2)) \, ,
\end{aligned}
\label{eq:resum_RGEs_general}
\end{equation}
where the anomalous dimensions $\gamma = \sum_i \left(\alpha_S(\mu^2)/\pi \right)^i \gamma^{(i)}$
are given as a power series in the strong coupling.
The $\mathcal{O}(\alpha_s)$ coefficients can be 
directly read off the NLO expressions explicitly given before, i.e.\
\begin{equation}
\gamma^{S_A(1)}_{a} = \frac{C_{a}}{2} \, , \, \, \, 
\gamma_{a}^{S_B(1)} = \frac{C_{a}}{2} \log \frac{1+r}{r}
\end{equation}
for the soft functions and
\begin{equation}
\gamma_q^{J_A(1)} = C_F \, , \, \, \, 
\gamma_q^{J_B(1)} = -\frac{3}{4} C_F , \, \, \, 
\gamma_g^{J_A(1)} = C_A \, , \, \, \, 
\gamma_g^{J_B(1)} = -\pi b_0 \, 
\end{equation}
for the jet functions. At NLL, the only 
necessary two-loop pieces are the $\mathcal{O}(\alpha_s^2)$ coefficients of
$\gamma^{S_A}_{a}$ and $\gamma^{J_A}_{q,g}$. They are related to the 
process-independent cusp-anomalous dimension according to
\begin{equation}
\gamma^{S_A}_{a} = \frac{C_{a}}{2} \gamma_\mathrm{cusp}\, , \, \, \, 
\gamma_q^{J_A} = C_F \gamma_\mathrm{cusp}\, , \, \, \, 
\gamma_g^{J_A} = C_A \gamma_\mathrm{cusp}\, , \, \, \, 
\end{equation}
where
\begin{equation}
\gamma^{(1)}_\mathrm{cusp}= 1 \, \quad \text{and}\quad
\gamma^{(2)}_\mathrm{cusp}= C_A \left(\frac{67}{36} - \frac{\pi^2}{12}\right) -\frac{5}{9} T_R N_f \, .
\end{equation}

The soft function in Mellin space is perturbatively well-behaved
for scales close to its natural scale, i.e.\ $\mu_s^2 \sim \hat{s}_T/\bar{N}^2$, 
because no large logarithms appear. 
Given that $Q^2 \sim \hat{s}_T$, we choose $Q^2/\bar{N}^2$ as the starting scale
for the renormalization-group evolution. With this choice, our results for
the exponentials derived below agree with \citere{deFlorian:2005fzc}.
The solution of the renormalization-group equation reads
\begin{equation}
\log \frac{S_a(N,\mu^2)}{S_a(N,Q^2/\bar{N}^2)} = - \int_{Q^2/\bar{N}^2}^{\mu^2} \frac{\text{d} \hat{\mu}^2}{\hat{\mu}^2} \left( \int_{\mu^2}^{\hat{\mu}^2} \frac{\text{d} q^2}{q^2} \gamma^{S_A}_{a}(\alpha_S(q^2)) + \gamma^{S_B}_{a}(\alpha_S(\hat{\mu}^2)) \right) .
\end{equation}
Separating and expanding the exponential containing only $\bar{N}$-independent terms 
to the desired order in
$\alpha_s$, one can cast the resummed soft function into the form
\begin{equation}
S_{a}(N,\mu^2) = \mathcal{S}_{a}(\mu^2) \, 
R^{\mathrm{exp}}(N,\mu^2,2\gamma^{S_A}_{a},2\gamma^{S_B}_{a},2) ,
\label{eq:resum_res_soft_final}
\end{equation}
where the exponential function for the resummed large logarithms reads	
\begin{equation}
\begin{aligned}
R^{\mathrm{exp}}(N,\mu^2,\gamma^{A},\gamma^{B},n) =
\exp \bigg[ -\int_{0}^{1-\frac{1}{\bar{N}}} \frac{\text{d} z}{1-z}
\bigg( &\int_{\mu^2}^{(1-z)^n Q^2} \frac{\text{d} q^2}{q^2} 
\gamma^{A}(\alpha_S(q^2))  \\
&+ \gamma^{B}\left(\alpha_S((1-z)^n Q^2)\right)
\bigg) \bigg]
\end{aligned}
\end{equation}
and one has
$\mathcal{S}_{a}(\mu^2) = S^{\mathrm{NLO}}_{a}(\mu^2/\bar{N}^2)$ 
at $\mathcal{O}(\alpha_s)$.

In analogy, the jet function in Mellin space
is perturbatively well-behaved for $\mu_c^2 \sim Q^2/\bar{N}$. We choose 
$Q^2/\bar{N}$ as the starting scale for the renormalization-group evolution. 
The resummed jet function is then given by
\begin{equation}
J_{a}(N,\mu^2) = \mathcal{J}_a(\mu^2) 
R^{\mathrm{exp}}(N,\mu^2,\gamma^{J_A}_{a},\gamma^{J_B}_{a},1)
\, , \label{eq:resum_jet_func_final} 
\end{equation}
where $\mathcal{J}_a(\mu^2) =J^{\mathrm{NLO}}_a(\mu^2/\bar{N})$ 
at $\mathcal{O}(\alpha_s)$.

To summarize, the NLL' result reads
\begin{equation}
\begin{aligned}
\text{d} \hat{\sigma}^{\mathrm{NLL'}}_{a}(N,\mu^2) = \text{d} \hat{\sigma}^{\mathrm{LO}}_a(N) \,\, C_{a}(\mu^2)\,\, 
& R^{\mathrm{exp}}(N+1,\mu^2,2\gamma^{S_A}_{a},2\gamma^{S_B}_{a},2)\\
& \times
R^{\mathrm{exp}}(N+1,\mu^2,\gamma^{J_A}_{a},\gamma^{J_B}_{a},1) \, ,
\end{aligned}
\end{equation}
where one needs $\gamma^{S_A}_{a}$, $\gamma^{J_A}_{a}$ up to $\mathcal{O}(\alpha_s^2)$ 
and $\gamma^{S_B}_{a}$, $\gamma^{J_B}_{a}$ up to $\mathcal{O}(\alpha_s)$. The matching coefficient
\begin{equation}
C_{a}(\mu^2) = \mathcal{S}_{a}(\mu^2) \mathcal{J}_{a}(\mu^2) H_{a}(\mu^2)
\end{equation}
can be truncated at $\mathcal{O}(\alpha_s)$. The $\mathcal{O}(\alpha_s)$ contributions to 
$C_{a}(\mu^2)$ due to $\mathcal{S}_{a}(\mu^2)$ and $\mathcal{J}_{a}(\mu^2)$ are
known analytically as discussed before, while the hard function is obtained
numerically with the help of MadLoop. The Mellin transform of the leading order cross section $\text{d}^2 \hat{\sigma}^{\mathrm{LO}}_a(N) / \text{d} p_T^2 \text{d} M^2$ is calculated analytically and can be expressed in terms of 
hypergeometric functions, see \refapp{app:LO}.

The resummed exponentials in Mellin space agree with the results in 
\citere{deFlorian:2005fzc}, where one can also find analytic expressions for 
the integrals in the exponential. 
As in \citere{deFlorian:2005fzc}, for our numerical
results, we resum large logarithms of $N$ instead of $\bar{N}$, i.e.\ 
$N$-independent exponentials containing $\gamma_E$ are expanded to 
$\mathcal{O}(\alpha_s)$ and lead to
additional terms proportional to $\gamma_E$ and $\gamma_E^2$ in 
$\mathcal{S}_{a}(\mu^2)$ and $\mathcal{J}_{a}(\mu^2)$.

\begin{figure}
\begin{center}
\includegraphics[width=10cm]{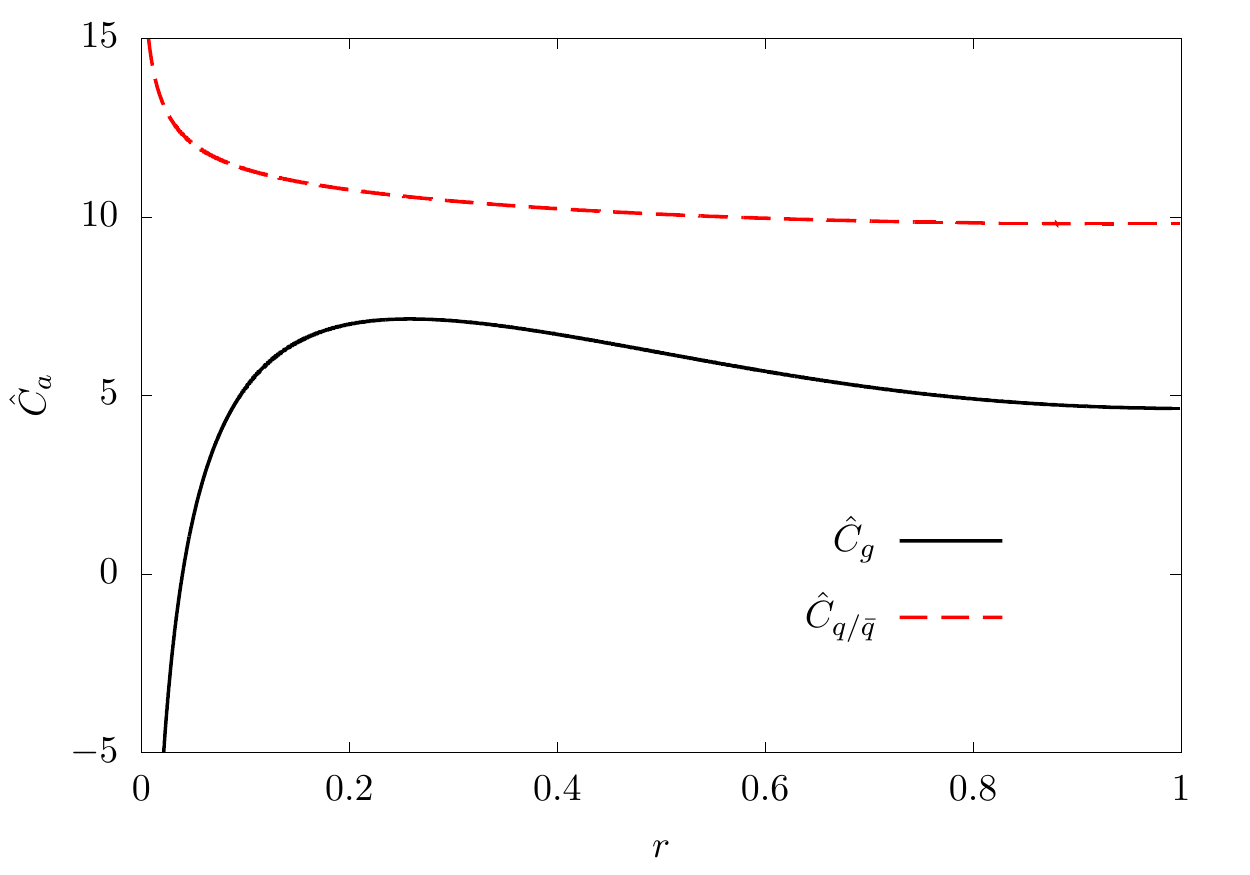}
\end{center}
\caption{\label{fig:matchingcoefficient} The scale independent part $\hat{C}_a$ 
of the matching coefficient as a function of $r=p_T/m_T$ for the two partonic channels 
as defined in \refeq{eq:matchingcoefficient}.}
\end{figure}

The scale $\mu$ is physically relevant because we define the PDFs
at $\mu_F$. Hence, we have to identify $\mu=\mu_F$. We have checked 
that the $\mu_F$-dependence of the matching coefficient is given by the known 
logarithms introduced via the PDF renormalization. In addition,
we have to define the numerical input for the strong coupling constant at some 
scale $\mu_R$. Within the
matching coefficient, all terms of $\mathcal{O}(\alpha_s)$ are simply evaluated 
using $\alpha_S(\mu_R)$. Hence the matching coefficient depends also on the choice of
$\mu_R$. The renormalization of the strong coupling in the LO cross section is part
of the hard function.
Also $R^{\mathrm{exp}}$ depends on the input value $\alpha_S(\mu_R)$ from which
$\alpha_S$ at another scale is derived using its renormalization-group 
running (see e.g.\ \citere{deFlorian:2005fzc}). 
The matching coefficient can be written as 
\begin{equation}
\label{eq:matchingcoefficient}
C_{a}(\mu_F^2,\mu_R^2) = \!\! \left( 1 + 
\frac{\alpha_S(\mu_R^2)}{\pi} 
\left(\hat C_{a} \!+ \!\left( \mathcal{P}_{a} - 2\, \gamma_E \gamma_{a}^{S_A (1)} - 
\gamma_E \gamma_a^{J_A (1)} \right) \log \frac{Q^2}{\mu_F^2} - 
\pi b_0 \log \frac{Q^2}{\mu_R^2} \right) \!\! \right) ,
\end{equation} 
where $\mathcal{P}_{g} = \frac{3}{2}C_F$ and 
$\mathcal{P}_{q} = \mathcal{P}_{\bar{q}}=\frac{3}{4}C_F + \pi b_0$. 
The constants $\mathcal{P}_{a}$ are related to PDF renormalization, where the $C_F$-pieces 
originate from the renormalization of the quark PDF and the piece proportional to $b_0$ 
from the renormalization of the gluon PDF. The terms proportional to $\gamma_E$ arise because we only keep $\log{N}$ instead of $\log{\bar{N}}$ in the exponentials, as mentioned above. The functions
$\hat C_{a}$ contain all scale-independent terms and include the hard contributions
which are only available numerically. 
The scale independent parts $\hat C_{a}$ of the matching coefficients only depend on the ratio $r=p_T/m_T$ and are presented in \reffig{fig:matchingcoefficient}.

To obtain numerical results from the resummed expression, we have to perform the inverse 
Mellin transform of the Mellin-space cross section, i.e.
\begin{equation}
\text{d} \sigma^{\mathrm{NLL'}}_{a}(s) = 
\frac{1}{2\pi i} \int_{c - i\infty}^{c + i \infty} \text{d} N \, \left(y_T^2\right)^{-N} 
\sum\limits_{i,j} f_i(N+1) \, f_j(N+1) \, \text{d} \hat{\sigma}^{\mathrm{NLL'}}_{a}(N) \, , 
\end{equation}
where $y_T=(p_T + m_T)/\sqrt{s}$ is the hadronic threshold variable and $f_i(N)$ are the Mellin moments of the PDFs.
Following the minimal prescription\cite{Catani:1996yz}, the constant $c$ is chosen to the left of the poles 
due to the running of the strong coupling $\alpha_s$ in the exponents of 
$\text{d}\hat{\sigma}^{\mathrm{NLL'}}_{a}(N)$, but to the right of all other poles of the integrand. 
To use the standard PDF implementations, which are not available in Mellin space, 
we rewrite the resummed hadronic cross section according to~\cite{Kulesza:2002rh}
\begin{equation}
\text{d}\sigma^{\mathrm{NLL'}}_{a}(s)= 
\sum\limits_{i,j} \int_0^1 \text{d} x_1 \, \tilde{f}_{i}(x_1) \int_0^1 
\text{d} x_2 \, \tilde{f}_{j}(x_2) \,
\text{d} \tilde{\sigma}^{\mathrm{NLL'}}_{a}(\hat{s})\, .
\end{equation}
Here, $\text{d} \tilde{\sigma}^{\mathrm{NLL'}}_{a}(\hat{s})$ is the inverse Mellin transform of 
$\text{d} \hat{\sigma}^{\mathrm{NLL'}}_{a}(N)/N^2$ with respect to the partonic threshold variable. Moreover, 
the $\tilde{f}_{i}$ are the inverse Mellin transforms of $N \,f_i(N+1)$, which are given by derivatives 
of the standard PDF functions according to
$\tilde{f}_{i}(x) = \frac{\text{d}}{\text{d}x}(-x f_i(x))$, and the sums run over all 
initial-state partons $i,j$ contributing to the LO process with parton $a$ in the final state. Introducing the convergence factor $1/N^2$ 
before performing the inverse Mellin transform of the resummed partonic cross section makes the numerical 
integration feasible.

\section{Numerical Results}
\label{sec:results}

In this section, we discuss numerical results for benchmark scenarios 
within the simplified model discussed in \refsec{sec:model}. 
We compute the full NLO predictions using MadGraph5\_aMC@NLO~\cite{Alwall:2014hca} using a 
FeynRules~\cite{Alloul:2013bka,Degrande:2014vpa}
implementation of the investigated dark-matter model which is publicly 
available~\cite{Backovic:2015soa,FeynRulesModelFile}.
The NLL' prediction which has been discussed in detail in \refsec{sec:resummed} is matched to this
NLO result. To avoid double counting, we add the NLO and the NLL' prediction and subtract the expansion of the 
resummed result to next-to-leading order in the strong coupling $\alpha_s$. Hence, only the contributions of
the resummed cross section beyond NLO are taken into account.

We use the PDF set PDF4LHC15\_nlo\_100 as provided by LHAPDF~\cite{Buckley:2014ana} throughout our analysis, i.e.\
for LO, NLO and NLO+NLL' predictions. Our central scale choice is given by
\[
\mu_F=\mu_R=\frac{H_T}{2}=\frac{m_T}{2}+\sum_j \frac{p_T^j}{2} \, ,
\]
where the sum runs over all final state jets. This is a common scale choice in fixed-order
calculations for processes involving jet production, cf.~\cite{Backovic:2015soa}.
Moreover, $H_T/2$ is a scale choice close to the two hard threshold variables 
$\sqrt{\hat{s}_T}$ and $Q$, since at threshold 
$\sqrt{\hat{s}_T}=m_T+p_T=H_T$ and $Q^2 = \frac{r}{1+r}H^2_T \le H^2_T/2$. We use $\alpha_s(\mu_R)$ as provided by LHAPDF
for the given PDF set. Moreover, we use $N_f=5$ as the number of light flavours. We present predictions
for the LHC at the center-of-mass energy $\sqrt{s}=13$~TeV. 
We use a fixed width for the mediator which is calculated at LO as discussed in \refapp{app:LO}.

The theoretical uncertainty of the NLO and the matched NLO+NLL' results due to missing higher-order corrections
is estimated in the usual way by independently varying the factorization scale $\mu_F$ and the 
renormalization scale $\mu_R$. To be precise, we use a standard 7-point variation, i.e.\ we make predictions
for 
\begin{equation}\label{eq:scale}
\mu_{F/R} = \xi_{F/R} \frac{H_T}{2} \, \, \, , \, \, \, (\xi_{F},\xi_{R}) = 
(1,1),\,(0.5,0.5),\,(2,2),\,(1,0.5),\,(1,2),\,(0.5,1),\,(2,1)
\end{equation} 
and show the minimal and maximal prediction as an error band around the nominal prediction
$(\xi_{F},\xi_{R}) = (1,1)$.

As discussed in \refsec{sec:model}, the vector and
axial-vector couplings of the mediator to the dark-matter particles and the standard-model quarks are set to
\begin{equation}
g^V_{\chi}=1 \, , \, g^A_{\chi}=0 \, , \, g^V_{SM} = 0.25 \, , \, g^A_{SM} =0 \, ,
\end{equation}
respectively. As can be seen in the analytic LO results presented in \refapp{app:LO}, the couplings 
to the SM quarks $g^{V,A}_{SM}$ enter the cross section explicitly as the sum of the 
squared vector and axial-vector couplings and in the mediator width. 
Hence, changing $g^{V,A}_{SM}$ mainly rescales the cross section by an overall factor which depends on 
whether the mediator is produced on-shell or off-shell. To a good approximation, the cross section also only depends on 
the squared sum of $g^{V,A}_{\chi}$ if the mediator mass is much larger than the dark-matter mass and the
production is dominated by mediators close to their mass shell. For $m_Y < m_\chi$, the production 
involves off-shell mediators and it is velocity suppressed near $M^2=4 m^2_\chi$ for axial-vector couplings. 
Hence, for axial-vector couplings the differential 
cross-section with respect to $M^2$ peaks at higher $M^2$, the cross section will be smaller 
compared to a model with a vector coupling of the same size but no qualitative differences are expected
for the radiative corrections.

The observable of interest is the missing transverse-momentum distribution. To obtain the $p_{T,\mathrm{miss}}$ spectrum,
the double-differential resummed result discussed in \refsec{sec:resummed} is numerically integrated with respect 
to the invariant mass $M^2$ of the dark-matter system.

At NLO precision, the model has been already investigated in \citere{Backovic:2015soa}. Using the setup of 
\cite{Backovic:2015soa}, we have reproduced the NLO cross section at the level of 0.5\% or better. 
For the LO results and the corresponding K-factors we have observed substantial differences. 
It turned out that the LO cross sections in~\citere{Backovic:2015soa} have not been obtained with the 
PDF set stated in the paper.

We discuss three benchmark points in detail. We have chosen two parameter sets for the mediator 
mass $m_Y$ and the dark-matter mass $m_\chi$ which lead to 
a large invariant mass of the dark-matter pair. In this case, the resummed cross section can be expected to reliably estimate 
the higher-order corrections beyond NLO accuracy. The first benchmark point is $(m_Y,m_{\chi})= (100,500)$~GeV, i.e.\ a 
relatively light mediator with a heavier dark-matter particle leading to off-shell production. The second 
benchmark point is $(m_Y,m_{\chi})= (1000,50)$~GeV, i.e.\ a heavy mediator with a relatively light dark-matter 
particle, which can be produced from an on-shell mediator. For these parameter points the resummed prediction 
expanded to NLO is slightly larger than the full NLO result and reproduces it within 5\% 
throughout the considered range for $p_{T,\mathrm{miss}}$. In addition, we investigate the model with the mass
combination $(m_Y,m_{\chi})= (100,1)$~GeV, i.e.\ a relatively light mediator with almost massless dark-matter
particles. This mass combination is kinematically very close to the main SM-background process, i.e.\
Z-boson production in association with a high-$p_T$ jet with subsequent invisible Z-boson decays into neutrinos. 
Since the final-state dark-matter system is
rather light in this case, the difference between the NLO-expanded resummed result and the full NLO prediction is
approximately -10\% of the NLO result at intermediate $p_{T,\mathrm{miss}}$ values around 500 GeV and is reduced to 
approximately -5\% at $p_{T,\mathrm{miss}}=2$~TeV.

For concreteness, we present integrated cross sections with a given cut on $p_{T,\mathrm{miss}}$ 
for our benchmark points in \reftab{tab:crosssection}. The LO, NLO, and matched NLO+NLL' results are shown along with
the corresponding K-factors.

\begin{table}[t]
\begin{center}
\def\arraystretch{1.3}
\begin{tabular}{|c|l|l|l|}
\hline
\begin{tabular}[c]{@{}c@{}}($m_Y,m_{\chi}$)\\ {[}GeV{]} \end{tabular}  &                                & $p_{T,\mathrm{miss}} >$ 500 GeV & $p_{T,\mathrm{miss}} >$ 1000 GeV  \\ \hline
\multirow{5}{*}{(100,500)}  & $\sigma_{\text{LO}}$ {[}pb{]}  & $5.18^{+1.11}_{-0.86} \times 10^{-4}$		&	$3.85^{+0.94}_{-0.71} \times 10^{-5}$		\\
                            & $\sigma_{\text{NLO}}$ {[}pb{]} & $6.96^{+0.43}_{-0.51} \times 10^{-4}$ 	& 	$5.15^{+0.32}_{-0.41} \times 10^{-5}$		\\
                            & $\sigma_{\text{NLO+NLL'}}$ {[}pb{]} & $7.25^{+0.22}_{-0.24} \times 10^{-4}$		&	$5.41^{+0.14}_{-0.16} \times 10^{-5}$		\\
                            & $\text{K}_{\text{NLO/LO}}$     & 1.34		& 1.34		\\
                            & $\text{K}_{\text{NLO+NLL'/LO}}$     & 1.40		& 1.40		\\ \hline
\multirow{5}{*}{(1000,50)}  & $\sigma_{\text{LO}}$ {[}pb{]}  & $6.65^{+1.37}_{-1.07} \times 10^{-2}$		& $3.85^{+0.92}_{-0.70}	\times 10^{-3}$	\\
                            & $\sigma_{\text{NLO}}$ {[}pb{]} & $8.74^{+0.48}_{-0.60} \times 10^{-2}$	& $5.11^{+0.30}_{-0.40} \times 10^{-3}$		\\
                            & $\sigma_{\text{NLO+NLL'}}$ {[}pb{]} & $9.03^{+0.26}_{-0.29} \times 10^{-2}$		& $5.34^{+0.13}_{-0.16} \times 10^{-3}$        \\
                            & $\text{K}_{\text{NLO/LO}}$     & 1.31		& 1.33		\\
                            & $\text{K}_{\text{NLO+NLL'/LO}}$     & 1.36		& 1.39		\\ \hline
\multirow{5}{*}{(100,1)}  & $\sigma_{\text{LO}}$ {[}pb{]}  & $7.26^{+1.36}_{-1.08} \times 10^{-1}$		&  $1.45^{+0.33}_{-0.25} \times 10^{-2}$		\\
                            & $\sigma_{\text{NLO}}$ {[}pb{]} & $1.08^{+0.09}_{-0.09} \times 10^{0}$	& $2.15^{+0.19}_{-0.20} \times 10^{-2}$		\\
                            & $\sigma_{\text{NLO+NLL'}}$ {[}pb{]} & $1.10^{+0.06}_{-0.05} \times 10^{0}$		& $2.22^{+0.10}_{-0.09} \times 10^{-2}$        \\
                            & $\text{K}_{\text{NLO/LO}}$     & 1.48		& 1.48		\\
                            & $\text{K}_{\text{NLO+NLL'/LO}}$     & 1.51		& 1.53		\\ \hline
\end{tabular}
\end{center}
\caption{\label{tab:crosssection} Integrated cross section for a given cut on $p_{T,\mathrm{miss}}$ for our benchmark
points at different levels of theoretical accuracy and the corresponding K-factors. The stated errors are 
determined by the 7-point scale variation defined in (\ref{eq:scale}).}
\end{table}

\begin{figure}[t]
\parbox{25cm}{
\includegraphics[width=8.2cm]{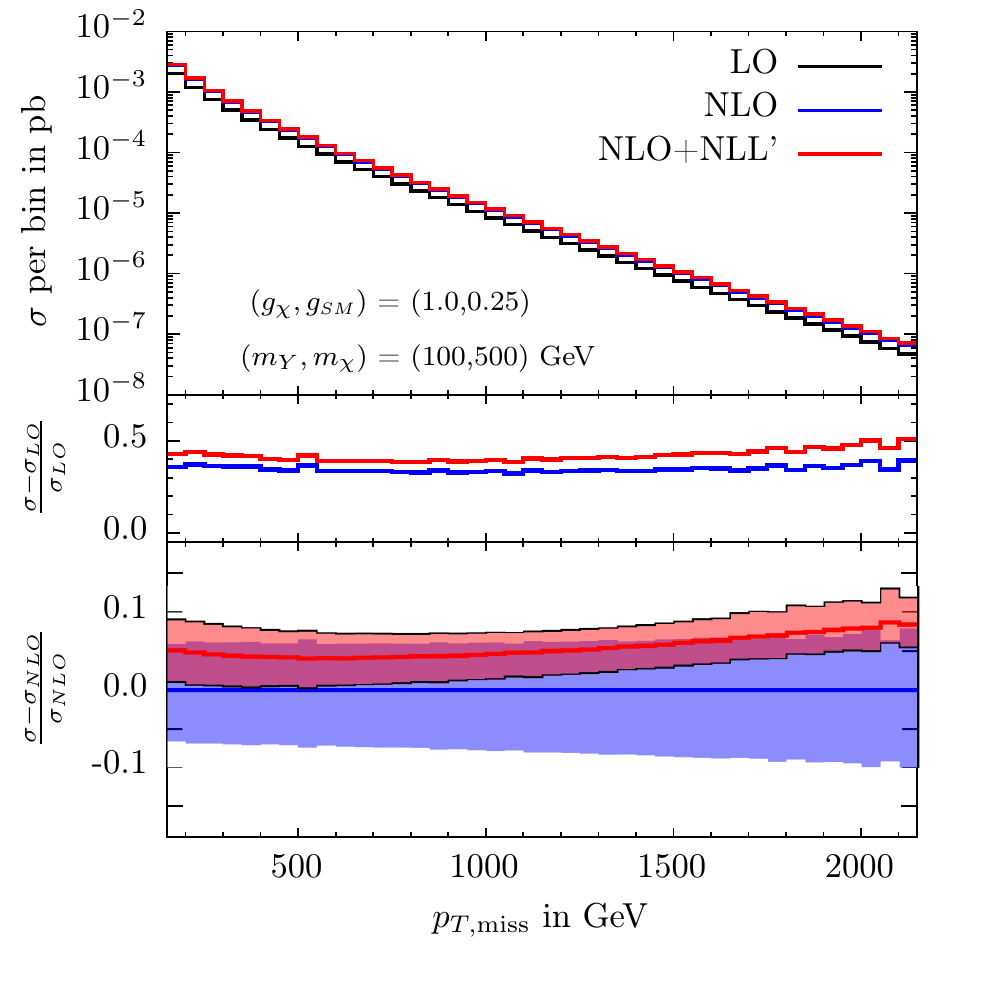}
\hspace{-.45cm}
\includegraphics[width=8.2cm]{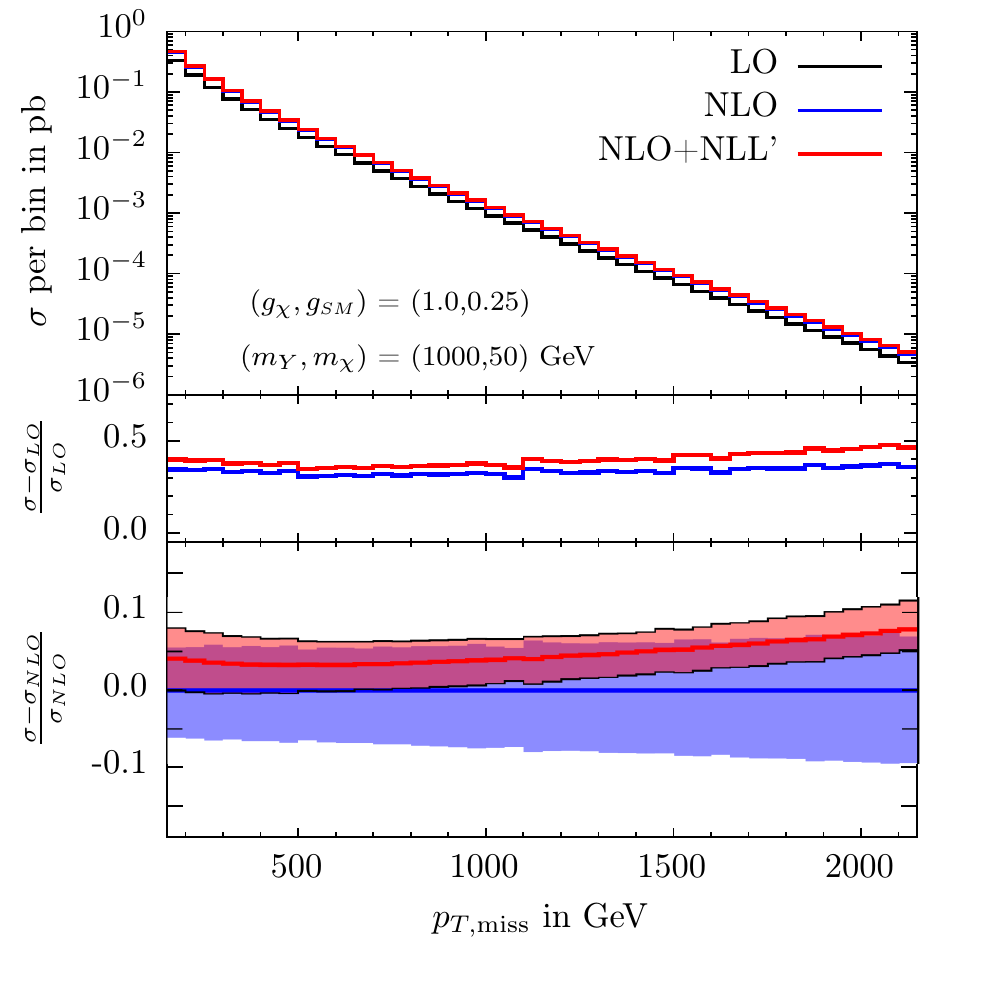}
}
\caption{\label{fig:pT} The missing transverse-momentum distribution for $(m_Y,m_{\chi})= (100,500)$ GeV (left) and 
$(m_Y,m_{\chi})= (1000,50)$ GeV (right) is shown in the upper panel at LO, NLO, and matched NLO+NLL' accuracy. 
In addition, the K-factors with respect to the LO prediction (middle panel) and the scale variation of the NLO and matched NLO+NLL' predictions (lower panel) are displayed. } 
\end{figure}

\begin{figure}[t]
\begin{center}
\includegraphics[width=8.2cm]{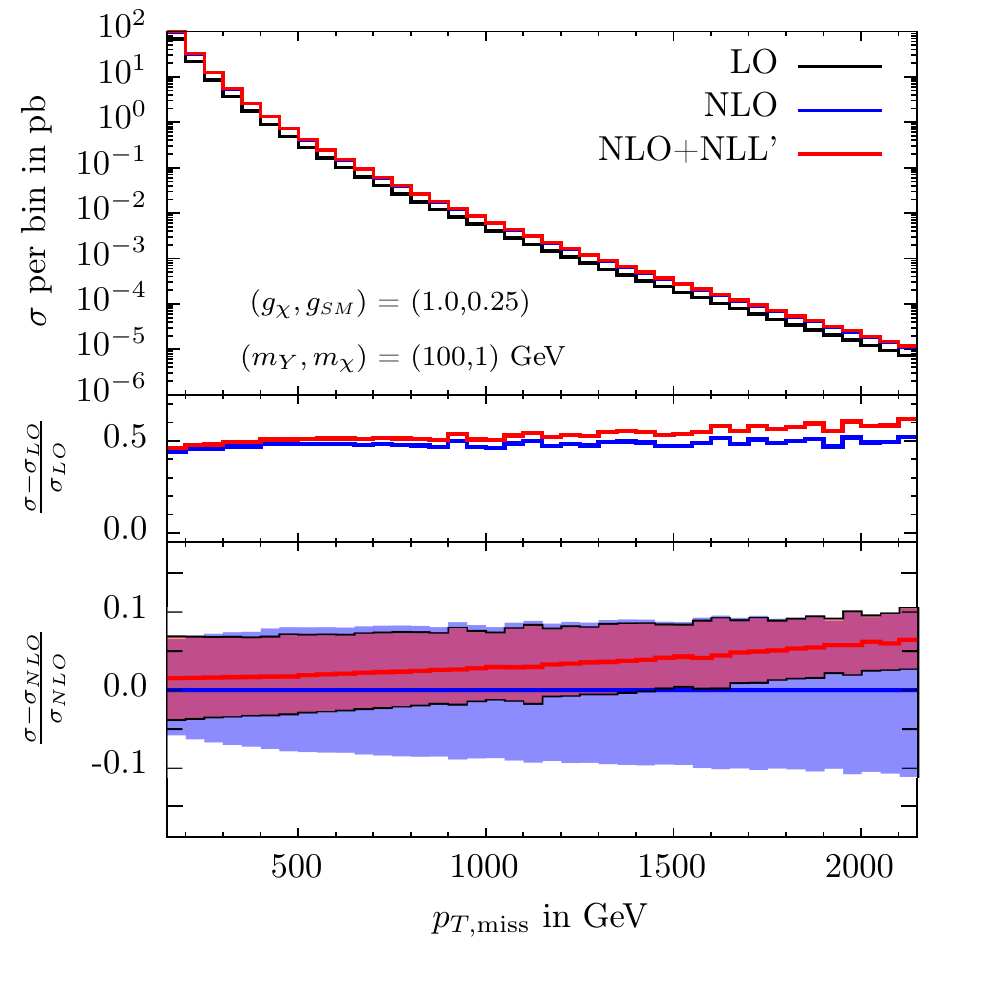}
\end{center}
\caption{\label{fig:pT_100_1} The missing transverse-momentum distribution for $(m_Y,m_{\chi})= (100,1)$ GeV
is shown in the upper panel at LO, NLO, and matched NLO+NLL' accuracy. 
In addition, the K-factors with respect to the LO prediction (middle panel) and the scale variation of the NLO and matched NLO+NLL' predictions (lower panel) are displayed. } 
\end{figure}

The
$p_{T,\mathrm{miss}}$ distributions are shown in \reffig{fig:pT} and \reffig{fig:pT_100_1}
in the range between 150~GeV and 2~TeV. 
The NLO K-factor for the heavy dark-matter systems in \reffig{fig:pT} does not vary much
as a function of $p_{T,\mathrm{miss}}$ and amounts to 30-40\%. For the lighter dark-matter 
system in \reffig{fig:pT_100_1} the NLO K-factor varies from 40 to 50\%. For mediator and dark-matter 
masses below 100~GeV, the K-factor grows and the accuracy of the expanded NLL' result is getting worse. 
For example, one finds $\text{K}\sim 1.7$ for $(m_Y,m_{\chi})= (10,1)$~GeV for a 
$p_{T,\mathrm{miss}}$ cut of 500~GeV. For smaller masses,
di-jet final-states start to sizeably  contribute, where the mediator is emitted from one of the jets. These
contributions cannot be expected to be well approximated by the threshold behavior of the mono-jet final-state.
Hence, in the following, we do not consider final states, for which the invariant mass of the dark-matter system 
is small.

For the benchmark points with a massive dark-matter final state shown in \reffig{fig:pT}, the
resummation adds roughly 5\% of the LO predictions on top of the NLO prediction in the region where the 
LHC is most sensitive. In the very tails of the distributions, the corrections due to resummation increases up to
12\% . Hence, as expected, the largest impact is observed
in the tail of the $p_{T,\mathrm{miss}}$-distribution at 2~TeV, where the dark-matter production takes place closer to
threshold. For $(m_Y,m_{\chi})= (100,1)$~GeV, the correction is almost negligible at small $p_{T,\mathrm{miss}}$ but
also rises to almost 10\% of the LO result in the tail of the distribution at 2~TeV. Note that it is essential 
to take into account the $\mathcal{O}(\alpha_s)$ terms of the matching coefficient in \refeq{eq:matchingcoefficient} 
on top of the resummed exponentials. Using NLL instead of NLL' results, the full NLO result 
is not approximated well by the expanded resummed result and the correction beyond NLO is much smaller.

The matched NLO+NLL' prediction also considerably reduces the scale uncertainty at large transverse momenta.
At large $p_{T,\mathrm{miss}}$ above 1~TeV the NLO error band is reduced by roughly 50\% for all three benchmark 
scenarios. For smaller values of $p_{T,\mathrm{miss}}$ the reduction of the size of the error band decreases with
decreasing $p_{T,\mathrm{miss}}$,
as shown in \reffig{fig:pT_100_1}. As expected, these results 
reproduce known results for vector-boson plus jet production in the Standard Model~\cite{Becher:2011fc}. Note that the
reduction of the scale error is less pronounced if we define the scale-variation error by a nine-point variation, i.e.\
also including $(\xi_{F},\xi_{R}) = (0.5,2),\, (2,0.5)$. The NLO error band is not affected by the different 
definition, while the width of the matched NLO+NLL' scale-variation band is increased at the upper end. 
Using the nine-point variation, there is no relevant improvement in the scale uncertainty for small transverse 
momenta. For $(m_Y,m_{\chi})= (100,1)$~GeV, the width of the matched NLO+NLL' error band is even bigger than the NLO
error band.
However, the error band of the fixed-order prediction grows to more than 15\% at transverse momenta close to 2~TeV, 
while the width of the error band for the matched predictions stays more or less constant at 10\%, using the nine-point
scale variation. 

\section{Conclusions}
\label{sec:Conclusions}

We have computed the resummed cross section for the hadroproduction of a dark-matter fermion pair in a simplified dark-matter model to NLO+NLL' accuracy. We have derived the 
corresponding soft and jet functions analytically using the strategy of regions. The hard function 
has been computed numerically using MadLoop. The resummation has been performed in Mellin-space, and we have
matched the resummed results to the full fixed-order NLO predictions obtained with MadGraph5\_aMC@NLO. 

We have presented numerical results for dark-matter production at the LHC for various 
mediator and dark-matter masses. For dark-matter pairs produced with a large invariant mass in the TeV range,
resummation increases the NLO prediction by approximately 5\% to 10\%. 
For smaller invariant masses of the dark-matter pair down to 100~GeV, the increase of the cross section beyond the 
NLO prediction is slightly reduced. For even smaller invariant masses, the threshold resummation as a means to 
estimate the leading higher-order corrections becomes questionable. Note that it is essential to 
include the NLO matching coefficient within the resummed result. We also observe a significant reduction in 
the scale uncertainty of the predictions, where the size 
of the NLO error band is reduced by 50\% at large transverse momentum. 

The combination of simple analytical 
formulas with numerical results from the MadGraph framework allows one to obtain the NLO+NLL' predictions for dark-matter production at the LHC in a straightforward manner. 

\section*{Acknowledgments}
This work has been supported by the German Research Foundation (DFG) through the Research Unit ``New Physics at the Large Hadron Collider'' (FOR 2239) and the CRC/Transregio ``P3H: Particle Physics Phenomenology after the Higgs Discovery'' (TRR 257). We are grateful to Kentarou Matawari for discussions and the comparison of numerical results. 

\begin{appendix}
\section{LO Cross Sections}
\label{app:LO}

To formulate the resummation in Mellin space, we need the Mellin moments of the
leading-order differential cross section for the two partonic channels. The
double-differential cross sections are given by
\begin{equation}
\frac{\text{d}^2 \hat{\sigma}_a^{\mathrm{LO}}}{\text{d}p_T^2 \text{d}M^2}  = 
\frac{2\alpha_s}{\pi} \,
\frac{\left(g^V_{SM}\right)^2 +\left(g^A_{SM}\right)^2}{m_T^2\, p_T^2} 
\, \frac{\hat{y}_T^2}{\sqrt{1-\hat{y}_T^2}}
\frac{M \, \Gamma^\chi_Y(M,m_\chi)}{(M^2-m_Y^2)^2 + \Gamma_Y^2 m_Y^2} 
\, f_a(\hat{y}_T^2,r) \, ,
\end{equation}
where
\begin{equation}
\Gamma^\chi_Y(M,m)= \frac{M}{12 \pi}\, \sqrt{1-\frac{4 m^2}{M^2}}\,
\left(  \left(g^V_{\chi}\right)^2 \left(1 + \frac{2m^2}{M^2}\right) + 
\left(g^{A}_{\chi} \right)^2 \left(1 - \frac{4m^2}{M^2}\right)
\right)
\end{equation}
results from the inclusive decay of the mediator and
\begin{equation}
\begin{aligned}
f_g(\hat{y}_T^2,r) & = \frac{2(r-1)^2 \hat{y}_T^4 -4 r^2 \hat{y}_T^2 + 2(r+1)^2}
{9 (r+1)^3 \sqrt{(r+1)^2 - \hat{y}_T^2 (r-1)^2}} \, , \\
f_q(\hat{y}_T^2,r) & = \frac{2 (r-1)^3 \hat{y}_T^6 + (r-2)(r-1)(r+2) \hat{y}_T^4 + 
                             (r+1)(4 r^2 - 3) \hat{y}_T^2 + (r+1)^3}
{24\,  (r+1)^4 \sqrt{(r+1)^2 - \hat{y}_T^2 (r-1)^2}}
\end{aligned}
\end{equation}
are functions which are specific for the given partonic production process.
Moreover, we use a fixed width in the propagator of the mediator given by
\begin{equation}
\Gamma_Y = \Gamma^\chi_Y(m_Y,m_\chi)\, \Theta\! \left(m_Y^2 - 4 m_{\chi}^2 \right)
+ N_C \Gamma^{\mathrm{SM}}_Y(m_Y,m_t)\, \Theta\! \left(m_Y^2 - 4 m_{t}^2 \right)
+  N_f N_C \Gamma^{\mathrm{SM}}_Y(m_Y,0) \, ,
\end{equation}
where the theta-functions determine which decay channels are
kinematically accessible for a given mediator mass, and
$\Gamma^{\mathrm{SM}}_Y$ is obtained from $\Gamma^\chi_Y$ by replacing the 
couplings to the dark-matter particle by the SM couplings to the quarks.
To be specific, we use $m_t=$172~GeV, and 
the width of the mediator for the benchmark points is given by 
$\Gamma_Y=(0.056,\, 0.025, \, 0.051) \, m_Y$ for $(m_Y,m_{\chi})= (1000,50),\, (100,500),\, (100,1)$~GeV,
respectively.

For the resummed
cross section, we need the Mellin-space representation of the cross section
according to 
\begin{equation}
\frac{\text{d}^2 \hat{\sigma}^{\mathrm{LO}}_{a}(N)}{\text{d} p_T^2 \text{d}M^2} =
\int_0^1 \text{d} \hat{y}_T^2 \, (\hat{y}^2_T)^{N-1} 
\frac{\text{d}^2 \hat{\sigma}^{\mathrm{LO}}_{a}(\hat{y}_T^2)}{\text{d} p_T^2 \text{d}M^2} \, .
\end{equation} 
Using
\begin{equation}
\int_0^1 \text{d} \hat{y}_T^2 \, \left(\hat{y}_T^2\right)^{N-1}  \frac{(\hat{y}_T^2)^n}
{\sqrt{1-\hat{y}_T^2} \sqrt{1- z \hat{y}_T^2}} = 
\frac{\sqrt{\pi} \, \, \Gamma(N+n) \,\,  _2F_1(\frac{1}{2},N+n;N +\frac{2n+1}{2};z)}{\Gamma(N+\frac{2n+1}{2})},
\end{equation} 
it can be expressed in terms of hypergeometric functions,
where $z=\frac{(r-1)^2}{(r+1)^2}$.

\end{appendix}

\providecommand{\href}[2]{#2}\begingroup\raggedright\endgroup

\end{document}